\documentclass[review]{elsarticle}

\usepackage{amsmath}

\newcommand{\refbr}[1]{(\ref{#1})}
\newcommand{\ov}{\overline}
\newcommand{\beq}{\begin{equation}}
\newcommand{\eeq}{\end{equation}}
\newcommand{\inI}[2]{= #1, \ldots, #2}

\begin{document}

\begin{frontmatter}
\title{Correction to the article "Dynamic
power management in energy-aware computer networks
and data intensive computing systems" published in "Future Generation Computer Systems" journal}
\author{Andrzej Karbowski}

\address{Institute of Control and Computation Engineering, Warsaw University of Technology, ul. Nowowiejska 15/19, 00-665 Warszawa, Poland,
E-mail:A.Karbowski@elka.pw.edu.pl}

\begin{abstract}
This paper indicates two errors in the formulation of the main
optimization model in the article "Dynamic power management in energy-aware computer
networks and data intensive computing systems" by Niewiadomska-Szynkiewicz et al. and shows how to fix them.
\end{abstract}


\end{frontmatter}

\section{Introduction - the original model}

The paper \cite{niewFGCS} presents an in-depth study  of the energy-aware traffic engineering in TCP/IP networks.
Authors have proposed therein an inspiring model of energy-aware router, an architecture of a control framework and various formulations of a network-wide energy saving optimization problem. They start from the exact mixed integer programming (MIP) formulation, which is aimed at solving the problem of a minimum energy routing. The objective is the minimization of the total power utilized by network components
 while ensuring end-to-end Quality of Service (QoS). The basic link-node formulation (\emph{LNPb}) is a network management problem
 with  binary decision variables describing full routing in a network and corresponding energy state assignments to all routers, line cards and communication ports.

 More precisely, the hierarchical network model proposed in \cite{niewFGCS} considers every  single communication
 port $p \in \{1,\ldots,P \}$ of every line card $c~\in~\{1,\ldots,C\}$ of a router  $r \in \{1,\ldots,R\}$.
The links connecting pairs of ports are denoted by $e \in \{1,\ldots,E\}$;
any network component can operate in $k \in \{1,\ldots,K\}$ energy state, but
two ports connected by a link are in the same state.
A demand $d \in \{1,\ldots,D\}$ is characterized by its source $s_d$, the destination $t_{d}$ port nodes and
the volume $V_d$.

The topology of the physical network is described by four matrices of binary indicators: $l_{cp}, g_{rc}, a_{ep}, b_{ep}$, whether, respectively: port $p$ belongs to the card $c$, card $c$ belongs to the router $r$,
link $e$ is outgoing from the port $p$ and link $e$ is incoming to the port $p$.

The decision variables are two vectors of binary indicators $x_{c}, z_{r}$ -
whether the card $c$ or router $r$ is used for data transmission and two incidence matrices
with elements: $y_{ek}$ - whether the link $e$ is in the state $k$ and $u_{ed}$ - whether the
demand $d$ uses the link $e$.

The full optimization problem presented in \cite{niewFGCS}  is as follows:
\begin{equation}
\min_{x_c,y_{ek},z_r,u_{ed}} \left[ F_{LNPb} = \sum_{r=1}^RT_{r}z_{r} + \sum_{c=1}^{C}W_{c}x_{c} + \sum_{e=1}^{E}\sum_{k=1}^{K}\xi_{ek}y_{ek} \right] ,
\label{funkcja}
\end{equation}
subject to the constraints:
\begin{align}
\forall_{\substack{e=1,\ldots,E}}& \quad \sum_{k=1}^{K}y_{ek} \leq 1,
\label{c1}\\
\forall_{\substack{d=1,\ldots,D,\\
	c=1,\ldots,C}}& \quad \sum_{p=1}^{P}l_{cp}\sum_{e=1}^{E}a_{ep}u_{ed} \leq x_{c},
\label{c2}\\
\forall_{\substack{d=1,\ldots,D,\\
	c=1,\ldots,C}}& \quad \sum_{p=1}^{P}l_{cp}\sum_{e=1}^{E}b_{ep}u_{ed} \leq x_{c},
\label{c3}\\
\forall_{\substack{r=1,\ldots,R,\\
	c=1,\ldots,C}}& \quad g_{rc}x_{c} \leq z_{r},
\label{c4}\\
\forall_{\substack{d=1,\ldots,D,\\
        r=1,\ldots,R,\\
	p=s_{d}}}&
           \sum_{c=1}^{C}
              g_{rc} l_{cp} \sum_{e=1}^{E}a_{ep}u_{ed}
         - \sum_{c=1}^{C}
              g_{rc} l_{cp} \sum_{e=1}^{E}b_{ep}u_{ed}
         = 1,
\label{c5}\\
\forall_{\substack{d=1,\ldots,D,\\
	r=1,\ldots,R\\
    p\neq t_d, p\neq s_d}}&
           \sum_{c=1}^{C}
              g_{rc} \sum_{p=1}^{P}
                 l_{cp} \sum_{e=1}^{E}a_{ep}u_{ed}
         - \sum_{c=1}^{C}
              g_{rc} \sum_{p=1}^{P}
                 l_{cp} \sum_{e=1}^{E}b_{ep}u_{ed}
         = 0,
\label{c6}\\
\forall_{\substack{d=1,\ldots,D,\\
        r=1,\ldots,R,\\
	p = t_{d}}}&
            \sum_{c=1}^{C}
              g_{rc} l_{cp} \sum_{e=1}^{E}a_{ep}u_{ed}
         - \sum_{c=1}^{C}
              g_{rc} l_{cp} \sum_{e=1}^{E}b_{ep}u_{ed}
         = -1,
\label{c7}\\
\forall_{\substack{e=1,\ldots,E}}& \quad \sum_{d=1}^{D}V_{d}u_{ed} \leq \sum_{k=1}^{K}M_{ek}y_{ek},
\label{c8}
\end{align}
where $M_{ek}$ and $\xi_{ek}$ are, respectively, the capacity and the power consumption of the link $e$ in the state $k$, and
$W_c$ and $T_r$ are power cost coefficients of  the card $c$ and the router $r$.

Unfortunately, there are some errors and deficiencies in this novel formulation.
They are pointed out in the next section and followed by suggestions how to fix them.
Without these changes the model does not describe well the dependencies between different
components of the energy-aware computer
networks and data intensive computing systems, including routers, cards, ports and links and is not fully useful.

\section{Errors in the original model and their correction}

There are two errors in the formulation \refbr{funkcja}-\refbr{c8}:
\begin{enumerate}
\item Flow conservation equations \refbr{c5}-\refbr{c7} are incorrectly written.\\
    First of all in Eqn. \refbr{c6} there is a redundancy in using the index p - at the beginning it is assumed fixed, while in the middle it is used as the index of a summation operator. It this equation, as well as in Eqs. \refbr{c5}, \refbr{c7},
    $p$ should not be a fixed parameter, because in computer networks ports
    are only labeled  inputs to routers  - nodes, where the switch of routes is done. Every port in a router
    can be an input or an output for signals and the summation over them and, at the same time,
    over all links outcoming and incoming to the router, should be performed in Eqs.
    \refbr{c5}, \refbr{c7} as it is in \refbr{c6}.\\

   \item Despite the announcement in subsection 4.1 of the article \cite{niewFGCS}: "Two ports
connected by the eth link are in the same state k.  "
there are no equations ensuring it. The conditions \refbr{c8} expressing the energy used by links are
formulated independently for all links. \\
The above assumption is natural in computers networks and it should be reflected in a good  model.
\end{enumerate}

To fix the two errors mentioned above it is proposed:
\begin{description}
\item[Ad.1.] To replace  the three equations \refbr{c5}-\refbr{c7}  by
        the following general one:
\beq
\forall_{\substack{d=1,\ldots,D,\\
        r=1,\ldots,R}}
            \sum_{c=1}^C g_{rc} \sum_{p=1}^P l_{cp} \sum_{e=1}^{E} \left(a_{ep}
         -   b_{ep} \right)u_{ed}
         =
\left\{\begin{array}{ll}
          1 & \,\; r=s_{d},\\
         -1 & \,\; r = t_{d},\\
          0 & \, \mbox{otherwise}.
         \end{array}\right.
\eeq
In this equation the summation is done across every router $r=1,\ldots,R$ for every demand $d=1,\ldots,D$.
All links connected to the router $r$ are taken into account owing to the
summations:
\beq
\sum_{c=1}^C \sum_{p=1}^P \sum_{e=1}^E g_{rc} l_{cp} a_{ep} u_{ed}
\eeq
for the outgoing traffic
and
\beq
\sum_{c=1}^C \sum_{p=1}^P \sum_{e=1}^E g_{rc} l_{cp} b_{ep} u_{ed}
\eeq
for the incoming traffic.
\item[Ad.2.] To augment the conditions \refbr{c8} with equality constraints assuring
    that the energy level in both links of every edge is the same.
    They are as follows:
    \beq
    \forall_{\substack{p = 1,\ldots,P\\
             k=1,\ldots,K}} \sum_{e=1}^E a_{ep} y_{ek} = \sum_{e=1}^E b_{ep} y_{ek}
             \label{dodat.rown}
    \eeq
    Since in equation \refbr{dodat.rown} indices $p$ and $k$ are fixed, with the assumptions
    taken, for a given port $p^*$ there is only one combination of links $e_1, e_2 \in 1,\ldots,E$, such that:
    \beq
    a_{e_1p^*}=b_{e_2p^*}=1, \forall_{e\neq e_1} a_{e,p^*}=0, \forall_{e \neq e_2} b_{e,p^*}=0.
    \eeq
    Taking this into account from equation \refbr{dodat.rown} we will get for all $k=1,\ldots,K$:
    \beq
    y_{e_1k}=y_{e_2k}
    \eeq
    The same reasoning may be repeated for the opposite port $p^{**}$ of the edge, such that:
    \beq
    b_{e_1p^{**}}=a_{e_2p^{**}}=1, \forall_{e\neq e_1} b_{e,p^{**}}=0, \forall_{e \neq e_2} a_{e,p^{**}}=0.
    \eeq
    It  means, that  the energy level will be the same in the edge formed of links $e_1$ and $e_2$.
\end{description}

The corrected model will have the form\footnote{We added
all sets of indices of the arguments of optimization and their domains, incorrectly omitted in \cite{niewFGCS}.}:
\begin{equation}
\min_{\substack{x_c,y_{ek},z_r,u_{ed}\\
c \in \ov{1,C}, e \in \ov{1,E}, k \in \ov{1,K}\\
r \in \ov{1,R}, d \in \ov{1,D}}} \left[ F_{LNPbC} =
\sum_{e=1}^{E}\sum_{k=1}^{K}\xi_{ek}y_{ek}
+ \sum_{c=1}^{C}W_{c}x_{c} +
\sum_{r=1}^RT_{r}z_{r}  \right] ,
\label{c1N}
\end{equation}
subject to the constraints:
\beq
\forall_{\substack{d=1,\ldots,D,\\
	c=1,\ldots,C}}  \quad \sum_{p=1}^{P}l_{cp}\sum_{e=1}^{E}a_{ep}u_{ed} \leq x_{c},
\label{c2N}
\eeq
\beq
\forall_{\substack{d=1,\ldots,D,\\
	c=1,\ldots,C}}  \quad \sum_{p=1}^{P}l_{cp}\sum_{e=1}^{E}b_{ep}u_{ed} \leq x_{c},
\label{c3N}
\eeq
\beq
\forall_{\substack{r=1,\ldots,R,\\
	c=1,\ldots,C}}  \quad g_{rc}x_{c} \leq z_{r},
\label{c4N}
\eeq
\beq
\forall_{\substack{e=1,\ldots,E}}  \quad \sum_{k=1}^{K}y_{ek} \leq 1,
\label{c5N}
\eeq
\beq
\forall_{\substack{d=1,\ldots,D,\\
        r=1,\ldots,R}}
            \sum_{c=1}^C g_{rc} \sum_{p=1}^P l_{cp} \sum_{e=1}^{E} \left(a_{ep}
         -   b_{ep} \right)u_{ed}
         =
\left\{\begin{array}{ll}
          1 & \,\; r=s_{d},\\
         -1 & \,\; r = t_{d},\\
          0 & \, \mbox{otherwise},
         \end{array}\right.
\label{c6-8N}
\eeq
\beq
\forall_{\substack{e=1,\ldots,E}}  \quad \sum_{d=1}^{D}V_{d}u_{ed} \leq \sum_{k=1}^{K}M_{ek}y_{ek},
\label{c9N}
\eeq
\beq
    \forall_{\substack{p = 1,\ldots,P\\
             k=1,\ldots,K}} \sum_{e=1}^E a_{ep} y_{ek} = \sum_{e=1}^E b_{ep} y_{ek},
    \label{rown.st.ener}
    \eeq

 \beq
x_c, z_r \in \{0,1\} \;\, c \inI{1}{C}; r \inI{1}{R};
\label{bina12}
\eeq
\beq
y_{ek},u_{ed} \in \{0,1\}\;\, e \inI{1}{E}; k \inI{1}{K}; d \inI{1}{D}.
\label{bina22}
\eeq

\section{Conclusions}

The paper fixes two errors in the basic formulation of the optimization problem
in the study \cite{niewFGCS}. To make the model correct it was necessary to modify the flow balance equations, treating routers as nodes (instead of ports as it is in \cite{niewFGCS}) and to add equality constrains
on the levels of power consumption in two links incoming to and outgoing from the same port.

\section*{Acknowledgments}
Before publication in Arxiv.org this paper was submitted to "Future Generation Computer Systems" journal
and rejected by the Editor-in-Chief with the words "Not withstanding the quality of your paper, I had to reject it. The reason being that FGCS receives a tremendous amount of manuscripts and we need to select the ones that are most urgent and of relevant importance to our readership. Your paper falls currently outside that scope and has not been forwarded to reviewers.".\\
It means that FGCS journal neglects a fundamental
in science right to critics.
Mistakes or errors are not so rare
in scientific publications and the proper approach is to
point them and, if possible, to fix them, e.g. \cite{IEEE-ToN-LL}, \cite{IEEE-ToN-ja}.
A journal editorial board cannot decline responsibility for the quality of the published material.


\end{document}